\documentstyle[12pt,epsf]{article}
\begin{document}
\setlength\textheight{8.75in}
\newcommand{\be}{\begin{equation}}
\newcommand{\ee}{\end{equation}}
\title{Dilatonic monopoles from ($4+1$)- dimensional vortices}
\author{{\large Yves Brihaye\footnote{ yves.brihaye@umh.ac.be}} \\
{\small Facult\'e des Sciences, Universit\'e de Mons-Hainaut, }\\
{\small B-7000 Mons, Belgium }\\
{ } \\
   {\large Betti Hartmann\footnote{ betti.hartmann@durham.ac.uk}}\\
{\small Department of Mathematical Sciences ,
University of Durham}\\
{\small Durham DH1 \ 3LE , United Kingdom}}

\date{\today}
%%%%%%\begin{titlepage}
\maketitle
\thispagestyle{empty}

\begin{abstract}
We study spherically and axially symmetric monopoles 
of the SU(2) Einstein-Yang-Mills-Higgs-dilaton (EYMHD) system
with a new coupling between the dilaton field and the covariant 
derivative of the Higgs field. This 
coupling arises in the study of ($4+1$)- dimensional vortices in the
Einstein-Yang-Mills (EYM) system.
\end{abstract}
\medskip
\medskip
\newpage
\section{Introduction}
In the 1920s, Kaluza and Klein studied the $5$-dimensional version of
the Einstein equations \cite{kaluza} by introducing a $5$-dimensional
metric tensor. When one dimension is compactified,
the equations of $4$-dimensional Einstein gravity plus 
Maxwell's equations are recovered. One of the new fields appearing in this model
is the dilaton, a scalar companion of the metric tensor.
In an analog way, this field arises in the
low energy effective action of superstring theories and is
associated with the classical scale invariance of these models \cite{maeda}.\

In recent years, a number of classical field theory models coupled to a dilaton
have been studied. It was found that if the solutions
of SU(2) Yang-Mills-Higgs (YMH) theory, namely the 't Hooft-Polyakov monopole
\cite{thooft} and its higher winding number generalisations \cite{ward,rebbi,fhp},
are coupled to a massless dilaton \cite{forgacs1,bh}, remarkable similarities to 
the qualitative features of Einstein-Yang-Mills-Higgs (EYMH) monopoles
\cite{weinberg, hkk2} arise. Especially, it was
observed that similarly to gravity the dilaton
can render attraction between like charged monopoles and thus
bound multimonopole states are possible. \

Recently, the YMH system  coupled to both
gravity and the dilaton has been
studied \cite{bhk,bh}. It was found that in the full 
Einstein-Yang-Mills-Higgs-dilaton (EYMHD) model, a simple relation between
the $tt$-component of the metric and the dilaton exists. The abelian
solutions to which the
configurations tend in the limit of critical coupling 
are the extremal Einstein-Maxwell-dilaton (EMD) solutions. These have
definite expressions for the energy and the value of the $rr$-component
of the metric at the origin which only depend on the fundamental couplings.\
      
Volkov argued recently \cite{volkov} that if $\frac{\partial}{\partial x_4}$
is a symmetry of the Einstein-Yang-Mills (EYM) system in ($4+1$) dimensions, where 
$x_4$ is the coordinate associated with the $5$th dimensions, then the
($4+1$)-dimensional EYM system reduces effectively to a ($3+1$)-dimensional EYMHD system
with a specific coupling between the dilaton field and the Higgs field.\

In this paper, we study both the spherically and axially symmetric solutions
of the ($3+1$)-dimensional EYMHD model deduced from the 
($4+1$)-dimensional EYM system. In Section 2, we give the EYMHD 
Lagrangian and review how the specific coupling arises. In Sections 3 und 4, 
we give the Ansatz and present our numerical results for the spherically and axially
symmetric solutions, respectively. In both these Sections emphasis
is placed on the flat space limit and on the limit
in which our system of equations reduces to the one in \cite{volkov}. 
The summary and conclusions are presented in Section 5.

\section{SU(2) Einstein-Yang-Mills-Higgs-dilaton theory}

The Lagrangian and the particular
coupling of the dilaton field $\Psi$ to the SU(2) gauge fields 
${A_{\mu}}^a$ and Higgs fields ${\Phi}^a$ ($a=1,2,3$), respectively,
arise effectively from the Lagrangian of ($4+1$)-dimensional
EYM theory. If both the matter functions and the metric functions
are independent on $x_4$, the $5$-dimensional fields can be
parametrized as follows (with $M,N=0,1,2,3,4$) \cite{volkov}:
\begin{equation}
g^{(5)}_{MN}dx^M dx^N = e^{-\zeta}g^{(4)}_{\mu\nu}dx^{\mu}dx^{\nu}-e^{2\zeta}
(dx^4)^2 \ , \ \mu , \nu=0, 1, 2, 3 
\end{equation}
and
\begin{equation}
A_M^{a}dx^M=A_{\mu}^a dx^{\mu}+\Phi^a dx^4  \  ,  \  \  a=1, 2, 3
\end{equation}
where $g^{(4)}$ is the $4$-dimensional metric tensor and $\zeta$ plays the
role of the dilaton.
Introducing a new coupling $\kappa$ to study the influence of the dilaton
systematically, we set $\zeta=2\kappa\Psi$ and obtain the following 
action of the effective $4$-dimensional EYMHD theory:
\begin{equation}
S=S_{G}+S_{M}=\int L_{G}\sqrt{-g^{(4)}}d^{4}x+ 
\int L_{M}\sqrt{-g^{(4)}}d^{4}x
\ . \label{action} \end{equation}
The gravity Lagrangian $L_{G}$ is given by
\begin{equation}
L_{G}=\frac{1}{16\pi G}R
\ , \end{equation}
where $G$ is Newton`s constant, while the matter 
Lagrangian $L_M$ reads:
\begin{equation}
L_{M}=-\frac{1}{4} e^{2\kappa\Psi}F_{\mu\nu}^{a}F^{\mu\nu,a}
-\frac{1}{2}\partial_{\mu}\Psi\partial^{\mu}\Psi
-\frac{1}{2}e^{-4\kappa\Psi}D_{\mu}\Phi^{a} 
D^{\mu}\Phi^{a}-e^{-2\kappa\Psi}V(\Phi^{a})
\ , \label{lag}  \end{equation}
with Higgs potential
\begin{equation}
V(\Phi^{a})=\frac{\lambda}{4}(\Phi^{a}\Phi^{a}-v^2)^2
\ , \end{equation}
the non-abelian field strength tensor
\begin{equation}
F_{\mu\nu}^{a}=\partial_{\mu}A_{\nu}^{a}-\partial_{\nu}A_{\mu}^{a}+
e\varepsilon_{abc}A_{\mu}^{b}A_{\nu}^{c}
\ , \end{equation}
and the covariant derivative of the Higgs field in the adjoint 
representation
\begin{equation}
D_{\mu}\Phi^{a}=\partial_{\mu}\Phi^{a}+
e\varepsilon_{abc}A_{\mu}^{b}\Phi^{c}
\ . \end{equation}
Here, $e$ denotes the gauge field coupling constant,
$\lambda$ the Higgs field coupling constant
and $v$ the vacuum expectation value of the Higgs field.

\section{Spherically symmetric solutions}

For the metric, the spherically symmetric Ansatz
in Schwarzschild-like coordinates reads \cite{weinberg}:
\begin{equation}
ds^{2}=g^{(4)}_{\mu\nu}dx^{\mu}dx^{\nu}=
-A^{2}(r)N(r)dt^2+N^{-1}(r)dr^2+r^2 d\theta^2+r^2\sin^2\theta
d^2\varphi
\label{metric}
\ , \end{equation}
with 
\begin{equation}
N(r)=1-\frac{2m(r)}{r}
\ . \end{equation}
In these coordinates, $m(\infty)$ denotes the (dimensionful) mass of
the field configuration.\

For the gauge and Higgs fields, we use the purely magnetic hedgehog ansatz
\cite{thooft}
\begin{equation}
{A_r}^a={A_t}^a=0
\ , \end{equation}
\begin{equation}
{A_{\theta}}^a= \frac{1-K(r)}{e} {e_{\varphi}}^a
\ , \ \ \ \ 
{A_{\varphi}}^a=- \frac{1-K(r)}{e}\sin\theta {e_{\theta}}^a
\ , \end{equation}
\begin{equation}
{\Phi}^a=v H(r) {e_r}^a
\ . \end{equation}
The dilaton is a scalar field depending only on $r$
\begin{equation}
\Psi=\Psi(r)
\ . \end{equation}
Inserting the Ansatz into the Lagrangian and varying with respect
to the matter fields yields the Euler-Lagrange equations,
while variation with respect to the metric yields the Einstein
equations.\

With the introduction of  dimensionless coordinates and fields
\begin{equation}
x=evr \ , \ \ \mu=evm \ ,\ \ \phi=\frac{\Phi}{v}\ , \ \ 
\psi=\frac{\Psi}{v}
\label{scale}
\ . \end{equation}
the Lagrangian and the resulting set of differential equations
depend only on three
dimensionless coup\-ling constants, $\alpha$, $\beta$ and $\gamma$,
\begin{equation}
\alpha =\sqrt{G}v =\frac{M_W}{eM_{\rm Pl}} \ , \ \
\beta=
 \frac{\sqrt{\lambda}}{e} = \frac{M_H}{\sqrt{2}M_W} \ , \ \
\gamma =\kappa v =\frac{\kappa M_W}{e}
\ , \label{coupling} \end{equation}
where $M_W=e v$, $M_H= \sqrt{2\lambda} v$ and $M_{\rm Pl}=1/\sqrt{G}$. With
the rescalings (\ref{scale}) and (\ref{coupling}), the dimensionless
mass of the solution is given by $\frac{\mu(\infty)}{\alpha^2}$.\

With (\ref{scale}) and (\ref{coupling}) the Euler-Lagrange equations read:
\begin{equation}
(e^{2\gamma\psi}ANK')'=A(e^{2\gamma\psi}\frac{K(K^2-1)}{x^2}+e^{-4\gamma\psi}H^2 
K)
\ , \label{dgl1} \end{equation}
\begin{equation}
(e^{-4\gamma\psi}x^2 ANH')'=AH(2e^{-4\gamma\psi}K^2+ \beta^2 x^2 
e^{-2\gamma\psi}(H^2-1))
\ , \label{dgl2} 
\end{equation}
\begin{eqnarray}
(x^2 AN\psi')' &=& 2\gamma A [e^{2\gamma\psi}(N(K')^2+\frac{(K^2-1)^2}{2 
x^2})
\nonumber \\
&-& e^{-2\gamma\psi}\frac{\beta^2 x^2}{4}(H^2-1)^2-2 
e^{-4\gamma\psi}(\frac{1}{2}
N (H')^2 x^2+H^2 K^2) ] 
\ , \label{dgl3} \end{eqnarray}
where the prime denotes the derivative with respect to $x$,
while we use the following combination of the Einstein equations
\begin{equation}
G_{tt}=2\alpha^2 T_{tt}=-2\alpha^2 A^2 N L_{M}
\ , \end{equation}
\begin{equation}
g^{xx}G_{xx}-g^{tt}G_{tt}=-4\alpha^2 N \frac{\partial L_{M}}{\partial N}
  \end{equation}
to obtain two differential equations for the two metric functions:
$$
%\begin{equation}
\mu ' = \alpha^2 \left(e^{2\gamma\psi}N(K')^2 + \frac{1}{2}N x^2(H')^2 
e^{-4\gamma\psi}+
\frac{1}{2x^2}(K^2-1)^{2} e^{2\gamma\psi}+K^2 H^2 e^{-4\gamma\psi}\right.
%\   \end{equation}
$$
\begin{equation}
 +  \left. \frac{\beta^{2}}{4}x^2
(H^2-1)^2 e^{-2\gamma\psi}+\frac{1}{2}Nx^{2}(\psi ')^2 \right)
\ , \label{dgl4} \end{equation}
\begin{equation}
A'=\alpha^2 x A \left(\frac{2(K')^2}{x^2}e^{2\gamma\psi}+
e^{-4\gamma\psi}(H')^2+(\psi ')^2
\right)
\ . \label{dgl5} \end{equation}
Since we are looking for globally regular, finite energy solutions which are
asymptotically flat, we impose the following set of boundary conditions:
\begin{equation}
K(0)=1 \ , \ \ H(0)=0 \ , \ \ \partial_{x}\psi|_{x=0}=0 \ , \ \ \mu(0)=0
\ . \label{bc1} \end{equation}
\begin{equation}
K(\infty)=0 \ , \ \ H(\infty)=1 \ , \ \ \psi(\infty)=0 \ , \ \ A(\infty)=1
\ , \label{bc2} \end{equation}
\subsection{Numerical results}
We have restricted our numerical calculations for the spherically as well
as for the axially symmetric solutions to $\beta=0$.
\subsubsection{The $\alpha=0$ limit}
For $\alpha=0$, the gravitational field equations (\ref{dgl4}) and (\ref{dgl5})
decouple from the rest of the system and we are left with the
YMHD model in flat space, $N(x)\equiv 1$ and $A(x)\equiv 1$.
In \cite{forgacs1} it was observed that (in analogy
to the EYMH system) the monopoles exist up to a maximal
value of the dilaton coupling $\gamma=\gamma_{max}$ and from there
on a second branch of solutions tend to the abelian solution 
for $\gamma\rightarrow\gamma_{cr} < \gamma_{max}$ with
$\psi(0)$ monotonically decreasing to $ -\infty$. Our numerical results 
indicate that no such $\gamma_{max}$ exists in the model studied here.
We have integrated the equations for $\gamma~\epsilon~[0:10]$
and found the solutions to exist for all these values of $\gamma$.
The profiles of the functions suggest that for $\gamma\rightarrow\infty$
the solutions tend to the vacuum solution with $K(x)\equiv 1$, $H(x)\equiv 0$
and $\psi(x)\equiv 0$. This is indicated by the fact that the mass
of the field configurations progressively tends to zero for rising $\gamma$ and
that the matter fields are equal to their vacuum values on increasing 
intervals of the coordinate $x$. To demonstrate this, we give below the
values of $x_K$ and $x_H$, where the gauge field $K(x)$ and the Higgs field $H(x)$, 
respectively, reach the value $0.5$, i.e. $K(x_K)=0.5$ and $H(x_H)=0.5$~:
\begin{center}
\begin{tabular}{|r|r|r|}
\hline
$\gamma$ & $x_K$ & $x_H$ \\
\hline
$0.1 $ & $2.2$  & $1.8$\\
$1.0 $ & $3.3$  & $2.1$\\
$5.0 $ & $11.2$ & $5.1$ \\
$10.0 $ & $22.4$ & $8.9$\\
\hline
\end{tabular}
\end{center}
Moreover, we find that $\psi(0)\geq 0$ for all $\gamma$. 
Since for $\gamma=0$, the BPS monopole solution is recovered, the curve
for $\psi(0)$ starts from zero at $\gamma=0$. From there it increases to a maximal
value at $\gamma\approx 1.38$ and then slowly decreases to zero 
for $\gamma\rightarrow \infty$. Together with (\ref{bc2}) our results
strongly suggest
that $\psi(x)$ tends to zero on the full interval $x~\epsilon~[0:\infty[$.\

The difference between the model studied here and the one studied in
\cite{forgacs1} is that in the standard case of the Yang-Mills-Higgs-dilaton
system, the equations are effectively the equations of an 
Einstein-Yang-Mills-Higgs model with metric
\begin{equation}
ds^{2}=-e^{2\kappa \Psi(r)}dt^2+e^{-2\kappa \Psi(r)}dr^2+
e^{-2\kappa \Psi(r)}r^2 d\theta^2+e^{-2\kappa \Psi(r)}r^2\sin^2\theta
d^2\varphi
\label{dilametric}
\end{equation} 
The components of the Einstein tensor then read:
\begin{equation}
G_{tt}=-\kappa e^{4\kappa \Psi(r)}\left[2\Psi^{''}
+4\frac{\Psi^{'}}{r}-\kappa (\Psi^{'})^2\right] \ , \ \ G_{rr}=-
\frac{G_{\theta\theta}}{r^2}=\frac{G_{\varphi\varphi}}{r^2\sin^2\theta}=
\kappa^2 (\Psi^{'})^2
\end{equation}
The following  combination of the Einstein equations
\begin{equation}
G_{t}^{t}-G_{r}^{r}-G_{\theta}^{\theta}-G_{\varphi}^{\varphi}=8\pi G (
T_{t}^{t}-T_{r}^{r}-T_{\theta}^{\theta}-T_{\varphi}^{\varphi})
\end{equation}
exactly gives the dilaton equation in the standard case with 
$\gamma^2=\kappa^2 v^2 \propto G v^2$. Equally, the Euler-Lagrange equations for the gauge and
Higgs field functions are obtained using the metric (\ref{dilametric}).
Thus the dilaton can be viewed as a metric field.
Since we know that
gravity leads to a critical value of the coupling constant, 
this is also true in the standard Yang-Mills-Higgs-dilaton case.
A horizon forms for $g_{tt}\rightarrow 0$
which implies $\Psi\rightarrow -\infty$. This is 
exactly the limiting solution found previously.\
 
Now, for the equations studied here, this is not possible. We cannot introduce a
$4$-dimensional metric which makes the equations in the limit of $A=N=1$ reduce
to a set of Einstein-Yang-Mills-equations. yThus, no critial $\gamma$
can be expected on the basis of the 
above argument. This is confirmed by our numerical results.\

The form of the Lagrangian (\ref{lag}) suggests that
for $\gamma$ getting bigger and
bigger, the only possibility to fulfill the requirement of finite 
energy is that
$\psi(x)\rightarrow 0$ on the full interval of $x$.

\subsubsection{The Volkov limit $\alpha^2=3\gamma^2$}
In \cite{volkov} only one fundamental coupling is given, the gravitational
coupling $G$. Comparing the $(4+1)$-dimensional EYM system
with the EYMHD system studied in this paper, we conclude that for
\begin{equation}
\alpha^2=3\gamma^2
\end{equation}
our system of equations reduces to the one in \cite{volkov}.\

Studying the full system of equations, we were particularly 
interested in reobtaining the results with a different numerical
method \cite{foot1}
and in studying some of the features of the solutions in greater detail.\
  
Volkov observed a spiraling behaviour of the parameters for the
EYM vortices. As expected and shown in Fig.~1, we observe this feature as well.
Our numerical results suggest that a number of branches 
exist, on which the minimum of the metric function $N$, $N_m$, and
the value of the metric function $A$ at the origin, $A(0)$, monotonically
decrease. In the limit of critical coupling,
the $rr$-component of the corresponding
$5$-dimensional metric tensor, expressed in $4$ dimensions
through the  function $g(x):=e^{2\gamma\psi}N(x)=
e^{\frac{2}{\sqrt{3}}\alpha\psi}N(x)$ develops a double zero at
a value $x_m > 0$ of the dimensionless coordinate $x$ \cite{volkov}. 
Our numerical results indicate that $x_m~\epsilon~ ]0:0.082]$.\

There exist several local maximal and minimal values of $\alpha$,
$\alpha_{max}^{(ij)}$ and $\alpha_{min}^{(ij)}$, respectively. By
$\alpha_{max,min}^{(ij)}$ we denote the value of $\alpha$, at which
the $i$-th and the $j$-th branch join. We find:
\begin{equation}
\alpha_{max}^{(12)}=1.268  \ , \ \  \alpha_{min}^{(23)}=0.312 \ , \ \ 
\alpha_{max}^{(34)}=0.419  \ , \ \  \alpha_{min}^{(45)}=0.395 \ 
\end{equation}
Apparantly, the difference between $\alpha_{max}$ and its corresponding
$\alpha_{min}$ decreases, i.e. the branches get smaller and smaller. 
We thus conjecture that the limiting solution 
is reached for a value of $\alpha$ 
close to $\alpha_{min}^{(45)}=0.395$.\

Progressing on the branches, the qualitative behaviour of the functions
changes \cite{volkov}. This is shown for the gauge field function $K(x)$ in Fig.~2.
While for all values of $\alpha$ on the first branch
(as an example, we show $K(x)$ for $\alpha=1.267$ close to 
$\alpha_{max}^{(12)}$)
and most of the values of $\alpha$ on the second branch (see the profile
for $\alpha=0.8$) $K(x)$ decreases
monotonically from its value at the origin $K(0)=1$, to its value at infinity
$K(\infty)=0$, oscillations of the gauge field start to occur on the second
branch for $\alpha\leq 0.4185$. For $\alpha_{min}^{(23)}=0.312$
the local minimum and maximum, respectively, are already quite pronounced, while
for $\alpha_{min}^{(45)}=0.395$, the location of both 
the minimum and maximum has moved to smaller values of $x$. We are convinced
that in analogy to \cite{volkov} the number of oscillations
increases when proceeding on further branches. In Fig.~3, we show
the value of the dilaton function $\psi$ at the origin (multiplied by $-1$),
$-\psi(0)$, and the mass of the solutions on the different branches
as functions of $\alpha$.
$\psi(0)$ is equal to zero for $\alpha=0$ (since this implies $\gamma=0$),
increases to a maximal value of $\psi(0)=0.174$ at $\alpha\approx 1.0$ and from there decreases
first to zero at $\alpha=1.220$ on the second branch and further decreases
to a minimal value of $\psi(0)=-5.274$ at $\alpha=0.314$ on the third branch.
From there it starts to increase to $\psi(0)=-4.847$ at $\alpha=0.395$
and then decreases again to $\psi(0)=-7.164$ at $\alpha_{min}^{(45)}=0.395$.\

The function $\psi(x)$ itself decreases monotonically from $\psi(0)$ to its
value $\psi(\infty)=0$ for all values of $\gamma$ on the first branch and for 
$\alpha\le 1.24$ on the second branch, staying positive for all values of $x$.
For $\alpha > 1.24$ on the second branch, a minimum (which has negative value)
starts to form and dips down deeper to negative values when progressing on the
branches. In the limit of critical coupling $\alpha_{cr}$, the dilaton
function $\psi_{EMD}$ of the corresponding EMD solution \cite{bhk}
\begin{equation} 
\psi_{EMD}=\frac{\sqrt{3}}{4\alpha_{cr}}\ln(1-\frac{X_{-}}{X}) \ , \ \ \
X_{-}=(\frac{4}{3})^{1/4}
\end{equation}
where the dimensionless coordinate $x$ is given in terms of $X$ \ , \
\begin{equation}
x=\alpha_{cr} X(1-\frac{X_{-}}{X})^{1/4}
\end{equation}
is reached for $x\geq x_m$, while it stays finite for $x~\epsilon ~[0:x_m[$. \

It has to be remarked here, though, that in the limit of critical
coupling, the gauge and Higgs field functions $K(x)$ and $H(x)$ don't reach
there abelian values $0$ and $1$ at $x=x_m$, respectively, but tend to the fixed point
described in \cite{volkov}. \

The mass of the solution stays close to one on the first branch and increases
monotonically on the second branch. 
On the third and fourth branch it differs only little from the
corresponding mass on the second branch and  thus the three different branches
can barely be  distinguished in Fig.~3. In the limit of critical
coupling $\alpha_{cr}$, 
the mass tends to the mass $\frac{\mu_{EMD}}{\alpha^2}$ of 
the corresponding EMD solution:
\begin{equation}
\frac{\mu}{\alpha^2} \rightarrow
\frac{\mu_{EMD}}{\alpha_{cr}^2}=\sqrt{\frac{3}{4}}\frac{1}{\alpha_{cr}}
\end{equation}

Let us finally investigate the Einstein-Yang-Mills-Higgs (EYMH) system
with the usual coupling of the dilaton  considered in \cite{bhk}
in the Volkov limit $\alpha^2=3\gamma^2$ . This model doesn't 
contain the prefactor of the covariant derivative involving the dilaton studied here.
In the BPS limit ($\beta=0$), the equation for the Higgs field
doesn't involve the dilaton field directly and vice versa,
while in the model studied here
it does even for $\beta=0$ (see (\ref{dgl2}) and (\ref{dgl3})). In Fig.~1, we show 
the values of $N_m$ and $A(0)$ for the model studied in \cite{bhk} for the
Volkov limit $\alpha^2=3\gamma^2$. The recalculation of the configurations
for this specific relation between $\alpha$ and $\gamma$ exactly confirms the results
obtained in \cite{bhk}. The solutions exist for $\alpha < \alpha_{max}=1.216$,
which fulfills the condition (48) of \cite{bhk}
\begin{equation}
\sqrt{\alpha_{max}^2+\gamma^2}=\sqrt{\frac{4}{3}}\alpha_{max}\approx 1.4 \ . \
\end{equation}
From there, on a second branch of solutions, the Einstein-Maxwell-dilaton (EMD)
solution is reached at $\alpha_{cr}$ with the minimum of the metric function
$N$, $N_m$, tending to the value $N(x=0)$ of the corresponding EMD solution
\cite{foot2}:
\begin{equation}
N_m \rightarrow N_{EMD}(0)=(\frac{\gamma^2}{\alpha^2+\gamma^2})^2=\frac{1}{16}
\end{equation} 
This is demonstrated in Fig.~1. $A(0)$ decreases to zero, while
$N_m > 0$, indicating that the extremal 
EMD solutions have a singularity at $x=0$ not hidden
by an horizon.

\section{Axially symmetric solutions}
Since the $n=1$ monopole was shown to be the unique spherically symmetric
solution in SU(2) YMH theory \cite{bogo}, we need to impose an axially symmetric
Ansatz (or one with even less symmetry) to construct higher winding number solutions. 
The axially symmetric Ansatz for the metric in isotropic coordinates reads:
\begin{equation}
ds^2=
  - f dt^2 +  \frac{m}{f} \left( d \tilde{r}^2+ \tilde{r}^2d\theta^2 \right)
           +  \frac{l}{f} \tilde{r}^2\sin^2\theta d\varphi^2
\ ,  \end{equation}
The functions  $f$, $m$ and $l$ now depend on $\tilde{r}$ and $\theta$. 
If $l=m$ and $f$ only depend on $\tilde{r}$, this metric reduces to the
spherically symmetric metric in isotropic coordinates and comparison
with the metric in (\ref{metric}) yields the coordinate transformation 
\cite{hkk1}:
\begin{equation}
\frac{d\tilde{r}}{\tilde{r}}=\frac{1}{\sqrt{N(r)}}\frac{dr}{r}
\end{equation}
For the gauge fields we choose the purely magnetic Ansatz \cite{rebbi}:
\begin{equation}
{A_t}^a=0 \ , \ \ \  {A_{\tilde{r}}}^a=\frac{H_1}{e\tilde{r}}{v_{\varphi}}^a 
\ , \end{equation}
\begin{equation}
{A_{\theta}}^a= \frac{1-H_2}{e} {v_{\varphi}}^a
\ , \ \ \ \ 
{A_{\varphi}}^a=- \frac{n}{e}\sin\theta \left(H_3{v_{\tilde{r}}}^a+
(1-H_4) {v_{\theta}}^a \right)
\ . \end{equation}
while for the Higgs field, the Ansatz reads \cite{rebbi,kkt}
\begin{equation}
{\Phi}^a=v (\Phi_1 {v_{\tilde{r}}}^a+\Phi_2 {v_{\theta}}^a)
\ . \end{equation}
The vectors $\vec{v}_{\tilde{r}}$,$\vec{v}_{\theta}$ and $\vec{v}_{\varphi}$
are given by:
\begin{eqnarray}
\vec{v}_{\tilde{r}}      &=& 
(\sin \theta \cos n \varphi, \sin \theta \sin n \varphi, \cos \theta)
\ , \nonumber \\
\vec{v}_{\theta} &=& 
(\cos \theta \cos n \varphi, \cos \theta \sin n \varphi,-\sin \theta)
\ , \nonumber \\
\vec{v}_{\varphi}   &=& (-\sin n \varphi, \cos n \varphi,0) 
\ .\label{rtp} \end{eqnarray} 
The dilaton field $\Psi$ now depends on $\tilde{r}$ and $\theta$ \cite{bh}:
\begin{equation}
\Psi=\Psi(\tilde{r},\theta)
\end{equation}
For $H_1=H_3=\Phi_2=0$, $H_2=H_4=K(\tilde{r})$, $\Phi_1=H(\tilde{r})$,
$\Psi=\Psi(\tilde{r})$ and $n=1$, this Ansatz reduces to the spherically symmetric Ansatz
in isotropic coordinates.\

The  Euler-Larange equations
arise by varying the Lagrangian with respect to the matter fields,
while we use the following combinations of the Einstein equations
\cite{kk2}
\begin{equation}
g_{\tilde{r}\tilde{r}}(G^{\mu}_{\mu}-2G^{t}_{t})=
16\pi G\frac{m}{f}(L_M+g^{tt}\frac{\partial L_M}{\partial g^{tt}}-
g^{\tilde{r}\tilde{r}}\frac{\partial L_M}{\partial g^{\tilde{r}\tilde{r}}}
-g^{\theta\theta}\frac{\partial L_M}{\partial g^{\theta\theta}}
-g^{\varphi\varphi}\frac{\partial L_M}{\partial g^{\varphi\varphi}})
\end{equation}
\begin{equation}
g_{\tilde{r}\tilde{r}}(G^{\tilde{r}}_{\tilde{r}}+G^{\varphi}_{\varphi})=
16\pi G\frac{m}{f}(L_M-g^{\tilde{r}\tilde{r}}
\frac{\partial L_M}{\partial g^{\tilde{r}\tilde{r}}}
-g^{\varphi\varphi}\frac{\partial L_M}{\partial g^{\varphi\varphi}})
\end{equation}
\begin{equation}
g_{\tilde{r}\tilde{r}}(G^{\tilde{r}}_{\tilde{r}}+G^{\theta}_{\theta})=
16\pi G\frac{m}{f}(L_M-g^{\tilde{r}\tilde{r}}\frac{\partial L_M}{\partial
 g^{\tilde{r}\tilde{r}}}
-g^{\theta\theta}\frac{\partial L_M}{\partial g^{\theta\theta}})
\end{equation}
to obtain $3$ differential equations for the metric functions $f$, $l$ and $m$,
which are diagonal with respect to the set of derivatives 
($f_{,\tilde{r},\tilde{r}}$,
$m_{,\tilde{r},\tilde{r}}$, $l_{,\tilde{r},\tilde{r}}$, 
$l_{,\theta,\theta}$, $l_{,\tilde{r},\theta}$).
With an analog rescaling as in (\ref{scale}), the
set of $10$ partial differential equations again 
depends only on the three fundamental coupling constants introduced in
(\ref{coupling}).\\
At the origin, the boundary conditions read (with $\partial_{\tilde{x}}=
\frac{1}{ev}\partial_{\tilde{r}}$):
\begin{equation}
\partial_{\tilde{x}}f(0,\theta)=\partial_{\tilde{x}}l(0,\theta)=
\partial_{\tilde{x}}m(0,\theta)=0,\ \ \partial_{\tilde{x}}\psi(0,\theta)=0
\end{equation}
\begin{equation}
H_i(0,\theta)=0,\ i=1,3 ,\ \ H_i(0,\theta)=1,\ i=2,4,\ \
\phi_i(0,\theta)=0,\ i=1,2 
\end{equation}
At infinity, the requirement for finite energy and asymptotically
flat solutions leads to the boundary conditions:
\begin{equation}
f(\infty,\theta)=l(\infty,\theta)=m(\infty,\theta)=1,\ \ \psi(\infty,\theta)=0
\end{equation}
\begin{equation}
H_i(\infty,\theta)=0,\ i=1,2,3,4 ,\ \ \phi_1(\infty,\theta)=1,\ \
\phi_2(\infty,\theta)=0 
\end{equation}
In addition,
boundary conditions on the symmetry axes (the $\rho$- and
$z$-axes) have to be fulfilled.
On both axes:
\begin{equation} 
H_1=H_3=\phi_2=0
\end{equation}
and
\begin{equation}
\partial_\theta f=\partial_\theta m=\partial_\theta l 
=\partial_\theta H_2=\partial_\theta H_4=
\partial_\theta \phi_1=\partial_\theta \psi=0
\end{equation}
\subsection{Numerical results}
Constructing the axially symmetric solutions numerically \cite{fidi}, we were mainly interested whether
bound multimonopoles are possible in this model and if so, how the strength
of attraction compares to that in \cite{bh}.
\subsubsection{The $\alpha=0$ limit} 
Since gravity itself is known to be attractive, we first studied the influence of
the dilaton alone. Constructing the $n=2$ solutions, we obtain
the same qualitative feature as for the $n=1$ case. The configurations
exist for all values of $\gamma$, tending to the vacuum solution
$H_1=H_3=\phi_1=\phi_2=\psi\equiv 0$, $H_2=H_4\equiv 1$ for $\gamma\rightarrow
\infty$. Again, we find that $\psi(0)\geq 0$ and that the curve
 for $\psi(0)$, starting from zero at $\gamma=0$, develops a maximum
at $\gamma=1.38$ (which coincides with the corresponding value for $n=1$). From
there it monotonically decreases to zero for $\gamma\rightarrow\infty$.

In Fig.~4, we show the difference between the mass
of the $n=1$ solution and the mass per winding number
of the $n=2$ solution $\Delta E=E(n=1)-E(n=2)/2$. The sign of
$\Delta E$ indicates  
whether like charged monopoles attract or repell each other, the modulus
of $\Delta E$ indicates the strength of attraction and repulsion, respectively. 
We find that for all values of $\gamma$ we have studied, the monopoles
are in an attractive phase and that with increasing $\gamma$ the strength of attraction
grows. For the model studied in \cite{bh}, the maximal $\Delta E$ 
was found to be $\Delta E_{max}\approx 0.0082$ at $\gamma=1.2$. From 
there $\Delta E$ decreases since the spherically as well as the axially symmetric
monopoles are tending to an essentially abelian, spherically symmetric solution
in the limit of critical coupling. In the model studied here, $\Delta E$
is smaller for the same values of $\gamma$ (e.g. we find that
$\Delta E (\gamma=1.2)\approx 0.0069$), but since no critical coupling exists,
$\Delta E$ grows with increasing $\gamma$. For e.g. $\gamma=5.0$,  we find
$\Delta E\approx 0.0167$, which is more that twice the maximal possible value
for the model in \cite{bh}. For large enough values
of $\gamma$, $\Delta E$ should decrease, because both the spherically and 
axially symmetric solution are tending to the vacuum. Since
the absolute value of the mass is decreasing to zero, the same
should happen for $\Delta E$. 
 
\subsubsection{The Volkov limit $\alpha^2=3\gamma^2$} 
Studying the $n=2$ axially symmetric monopoles, we find the same
qualitative features as in the spherically symmetric case. We obtain:
\begin{equation}
\alpha_{max}^{(12)}=1.279     \ , \ \ \alpha_{min}^{(23)}=0.295
\end{equation}
These values are slightly bigger and smaller, respectively, than
the corresponding values for the $n=1$ solution. It has already been
observed previously, that the branches for the $n=2$ solutions exist
for higher values of the maximal coupling \cite{hkk1}. We find here that
an analog  holds true for the minimal coupling, namely that the $n=2$ solutions
exist for smaller minimal values than in the $n=1$ case. \

The behaviour of $\psi(0)$ is completely analog to the $n=1$ case~: starting from
zero at $\alpha=0$, it increases to a maximal value of $\psi(0)\approx 0.178$ at 
$\alpha\approx 1.0$, then decreases to zero at $\alpha\approx 1.19$
and reaches $\psi(0)\approx -5.324$ at $\alpha_{min}^{(23)}=0.295$.
The dilaton function, which has only a weak angle-dependence, is positive
and monotonically decreasing on the first branch and 
starts to form a negative valued minimum on the second branch.
Proceeding on the second branch, the solutions are less and less angle dependent
and the maximum of the function $H_1$, $H_3$ and $\Phi_2$ decreases.  
We didn't manage to construct further branches with our numerical routine 
(since these branches are rather small), but believe that in analogy to the $n=1$ solutions,
the $n=2$ solutions develop a double zero, which  is now 
(because of the choice of isotropic coordinates) located at $\tilde{x}=0$.\

Comparing $\Delta E$ for the model studied here and for that in \cite{bh}, 
we find that the different values don't deviate very much from each other.
However, in the model studied in \cite{bh} the parameter space, in which
solutions exist, is limited by $\sqrt{\alpha^2+\gamma^2}\approx 1.4$.
This means that
in the Volkov limit solutions only exist for $\alpha\leq \alpha_{max}\approx 1.21$.
Here, the solutions exist for bigger values of $\alpha=\sqrt{3}\gamma$ and thus
the maximum of $\Delta E$ is slightly bigger with $\Delta E_{max}\approx 0.011$.
This is reached on the first branch of solutions as demonstrated in Fig.~4.
The second branch starts from a value of $\Delta E$ well below the corresponding
one on the first branch and from there decreases. For all values of
$\gamma=\frac{\alpha}{\sqrt{3}}$, the value of $\Delta E$ on the first branch exceeds the value of
$\Delta E$ on the second branch.  

\section{Conclusion and Summary}   
We have studied spherically and axially symmetric solutions of SU(2) EYMHD
theory. The specific coupling between the dilaton and the Higgs field
in this model arises from a ($4+1$)-dimensional EYM model studied recently in
\cite{volkov}. We find that in the flat space limit ($\alpha=0$), the solutions
exist for all values of $\gamma$ tending to the vacuum solution
for $\gamma\rightarrow \infty$. This contrasts to the situation in the
usual YMHD model studied in \cite{forgacs1}, where the solutions only exist
for $\gamma\leq\gamma_{max}$ tending to an abelian solution in the limit of
critical coupling $\gamma_{cr}$. In both models, the dilatonic monopoles
are in an attractive phase. Since no critical coupling exists in the model studied
here, much stronger bound monopoles are possible.\

Apparently, unlike the YMHD model with the usual coupling,
the YMHD system studied here doesn't have
analog features than the EYMH system. Moreover, the relation
between the $tt$-component of the metric and the dilaton found
in \cite{bhk}, doesn't exist

When in addition gravity is coupled, our equations agree with those in \cite{volkov}
for a specific relation between the gravitational coupling $\alpha$ and the
dilaton coupling $\gamma$, namely $\alpha^2=3\gamma^2$. Studying the spherically and 
axially symmetric solutions of these equations in this limit, we observe the same
spiraling behaviour of the parameters found in \cite{volkov}. Several branches exist
differing from the situation in the model with the usual dilaton coupling studied
in \cite{bhk}, where maximally two branches exist. Like in \cite{bh}
the dilatonic monopoles can form bound multimonopole states.\

In \cite{bhk}, non-abelian black holes of the EYMHD model were studied 
giving further evidence that the "No-hair" conjecture doesn't hold
in models involving non-abelian fields.
It would be interesting to construct the analogs in this system, especially
the axially symmetric non-abelian black holes. Their existence would
show that Israel's theorem cannot be extended to EYMHD theory, like it has
previously been observed in EYMH theory \cite{hkk1}.\

Finally, since in the low energy effective action of string theory, 
further corrections to the equations of standard physics, such as higher
order corrections to gravity in form of Gauss-Bonnet terms or new 
fields like the axion field arise, it would be interesting to study their influence
on the solutions constructed here.\\  

{\bf Acknowledgements}
B. H. was supported by the EPSRC.

\newpage

\begin{figure}\centering\epsfysize=15cm
\mbox{\epsffile{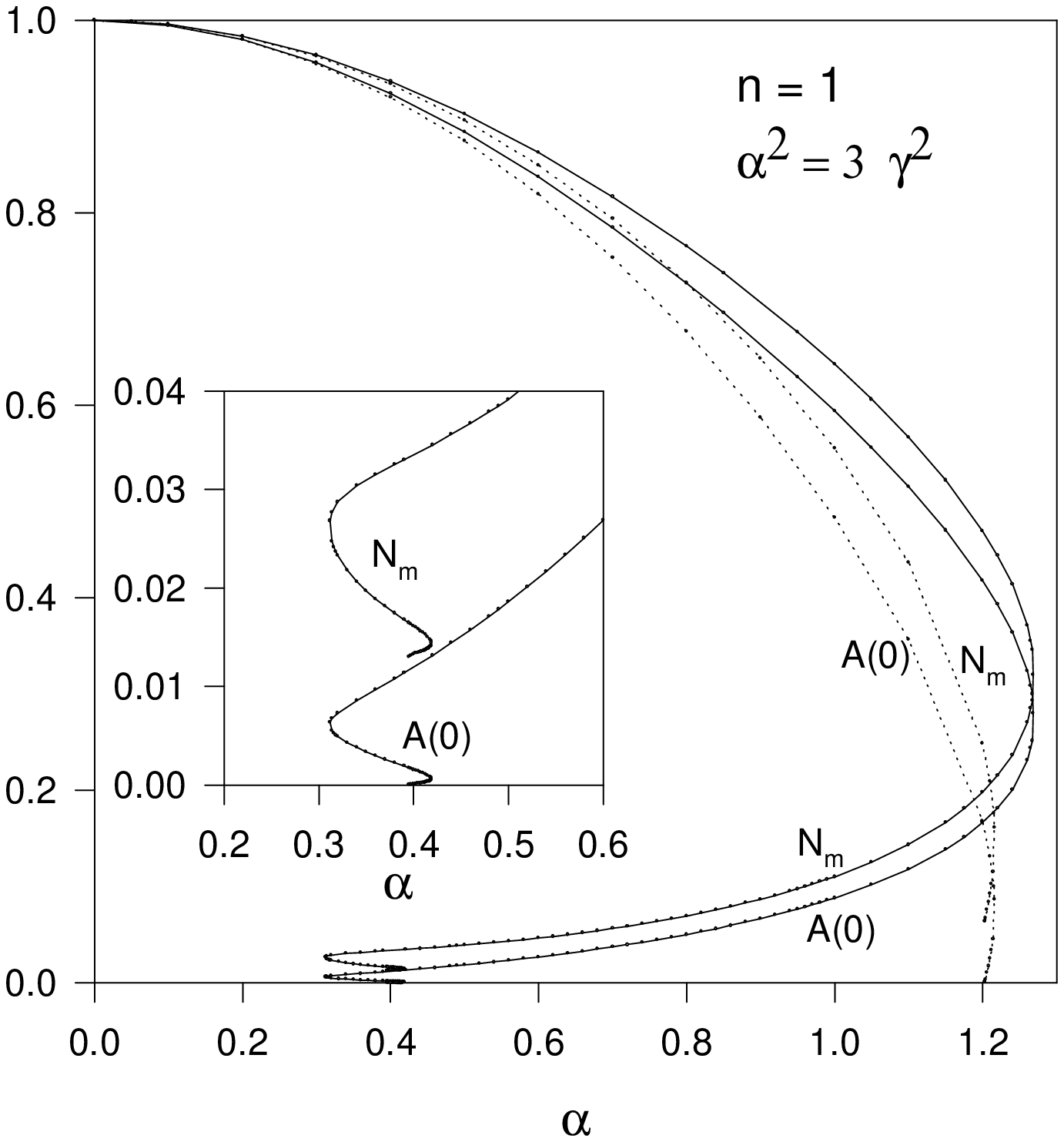}}
\caption{The values of the minimum of the metric function $N$, $N_m$, and
of the metric function $A$ at the origin, $A(0)$, are shown as a function
of $\alpha$ for the $n=1$ solutions in the Volkov limit $\alpha^2=3\gamma^2$. The solid 
 and dashed lined curves denote the values obtained in the model
studied here and the one studied in \cite{bhk}, respectively.}
\end{figure}
\newpage
\begin{figure}\centering\epsfysize=15cm
\mbox{\epsffile{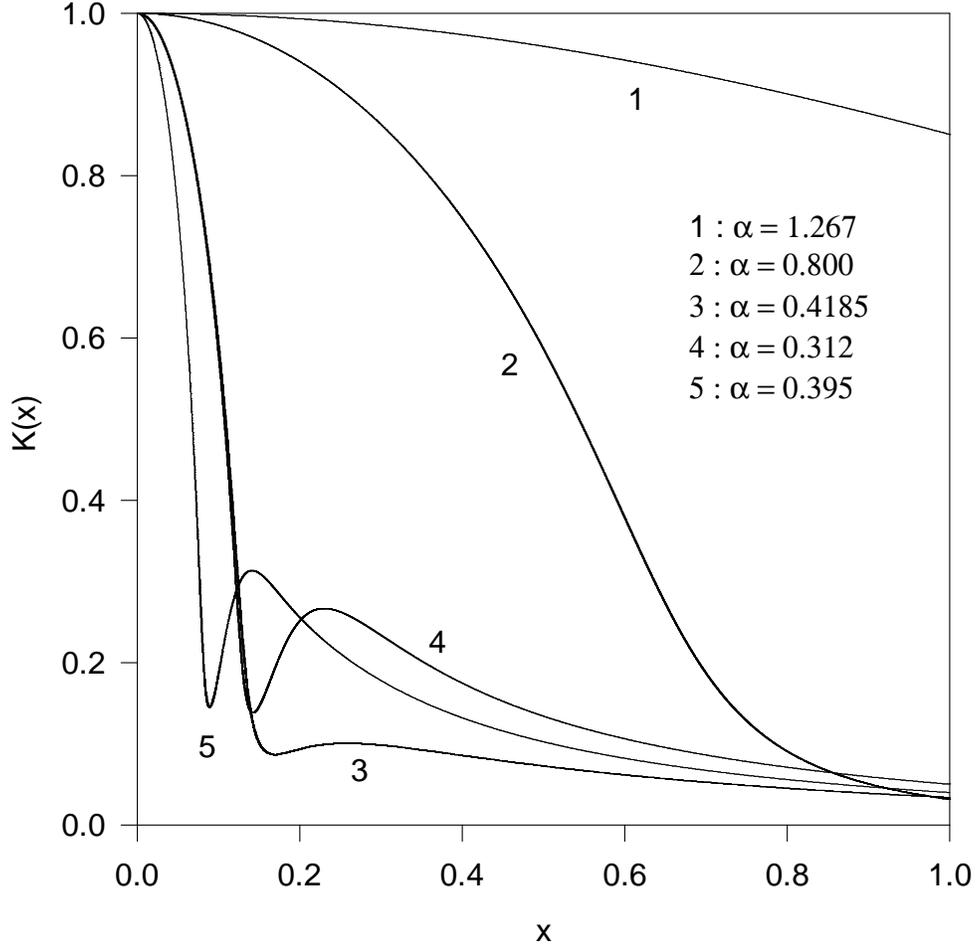}}
\caption{The gauge field function $K(x)$ is shown as function of the
dimensionless coordinate $x$ for the $n=1$ solutions in the Volkov limit 
$\alpha^2=3\gamma^2$
for the following five values of $\alpha$~: 
$1:\alpha=1.267$ close to $\alpha_{max}^{(12)}$ , $2:\alpha=0.8$ (second branch),
$3:\alpha=0.4185$ (second branch), 
$4:\alpha_{min}^{(23)}=0.312$ and $5:\alpha_{min}^{(45)}=0.395$.
}
\end{figure}

\newpage
\begin{figure}\centering\epsfysize=15cm
\mbox{\epsffile{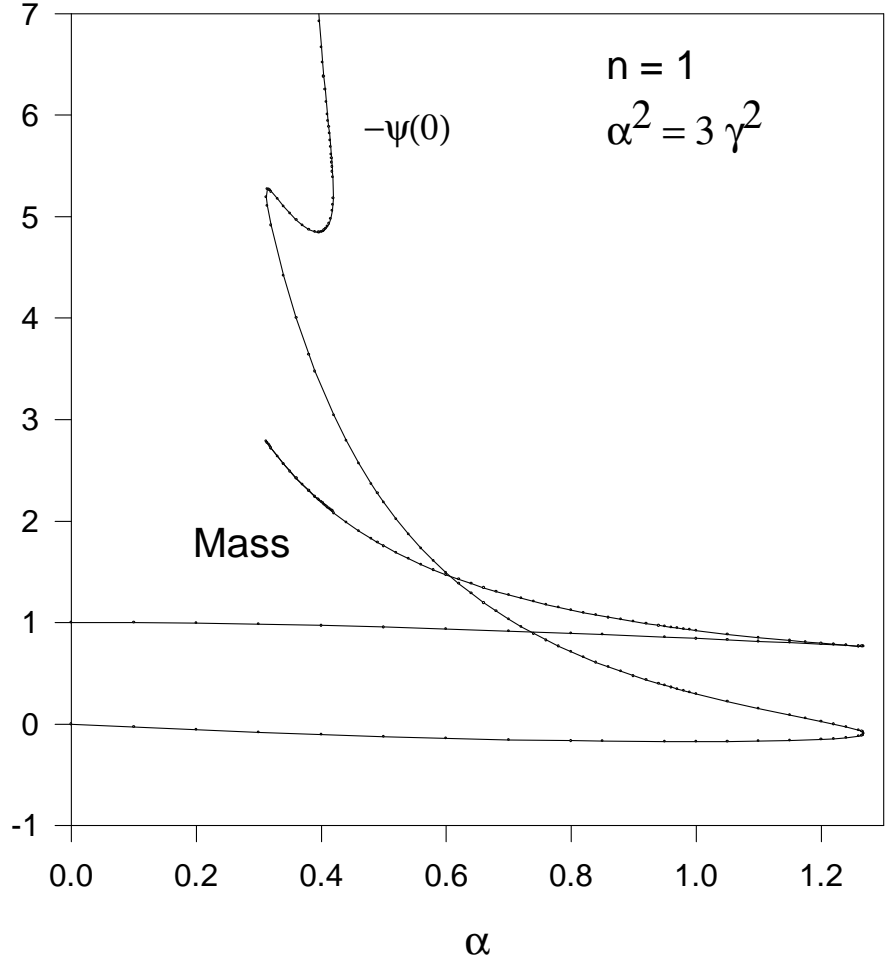}}
\caption{The value of the dilaton function $\psi$ at the origin (multiplied by $-1$),
-$\psi(0)$, is shown as function of $\alpha$ for the $n=1$ solutions in the Volkov limit
$\alpha^2=3\gamma^2$. Also shown is the mass of these solutions. }
\end{figure}

\newpage
\begin{figure}\centering\epsfysize=15cm
\mbox{\epsffile{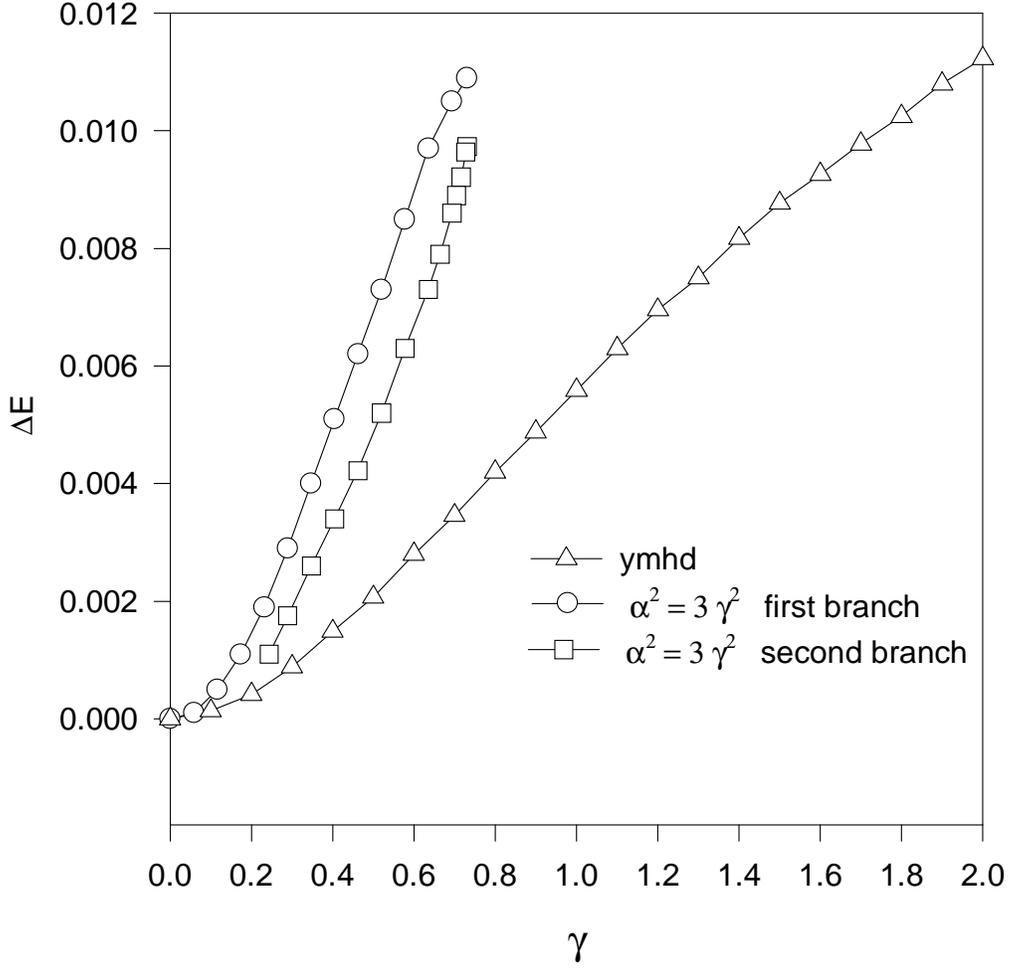}}
\caption{The difference between the mass
of the $n=1$ solution and the mass per winding number
of the $n=2$ solution $\Delta E=E(n=1)-E(n=2)/2$ is shown as function of $\gamma$ for
the first (circles) and second (squares) branch arising in the EYMHD system in the Volkov limit
$\alpha^2=3\gamma^2$. 
Also shown is $\Delta E$ for the YMHD solutions (triangles). Note that for
the EYMHD solutions, the $\gamma$-axis represents a rescaled $\alpha$-axis.}
\end{figure}


\begin{thebibliography}{99}
\bibitem{foot1}
To integrate the equations, we used the differential equation
solver COLSYS which involves a Newton-Raphson method \cite{colsys}.
\bibitem{foot2}
In \cite{bhk}, eq.~(45) contains an error, since an overall square
is missing on the rhs of that equation.
\bibitem{fidi}
To solve the set of partial differential equations, we used the 
numerical routine used and described e.g. in \cite{hkk1}.
\end{thebibliography}
\end{document}